\documentclass[aps,prb,twocolumn,byrevtex]{revtex4}

\usepackage{times,mathptm}
\usepackage[dvips]{graphicx,color}

\begin{document}

\title{Microscopic mechanism of tunable band gap in potassium doped few-layer black phosphorus}
\author{Sun-Woo Kim,$^{1}$ Hyun Jung,$^{1}$ Hyun-Jung Kim,$^{2}$ Jin-Ho Choi,$^{3,1}$ Su-Huai Wei,$^{4,{\dagger}}$ and Jun-Hyung Cho$^{1,*}$}
\affiliation{$^1$ Department of Physics and Research Institute for Natural Sciences, Hanyang University,
222 Wangsimni-ro, Seongdong-gu, Seoul, 133-791, Korea \\
$^2$ Korea Institute for Advanced Study, 85 Hoegiro, Dongdaemun-gu, Seoul 130-722, Korea \\
$^3$ Research Institute of Mechanical Technology, Pusan National University, 2 Pusandaehak-ro 63 beon-gil, Geumjeoung-gu, Korea\\
$^4$ Beijing Computational Science Research Center, Beijing 100094, China}
\date{\today}

\begin{abstract}
Tuning band gaps in two-dimensional (2D) materials is of great interest in the fundamental and practical aspects of contemporary material sciences. Recently, black phosphorus (BP) consisting of stacked layers of phosphorene was experimentally observed to show a widely tunable band gap by means of the deposition of potassium (K) atoms on the surface, thereby allowing great flexibility in design and optimization of electronic and optoelectronic devices. Here, based on the density-functional theory calculations, we demonstrates that the donated electrons from K dopants are mostly localized at the topmost BP layer and such a surface charging efficiently screens the K ion potential. It is found that, as the K doping increases, the extreme surface charging and its screening of K atoms shift the conduction bands down in energy, i.e., towards higher binding energy, because they have more charge near the surface, while it has little influence on the valence bands having more charge in the deeper layers. This result provides a different explanation for the observed tunable band gap compared to the previously proposed giant Stark effect where a vertical electric field from the positively ionized K overlayer to the negatively charged BP layers shifts the conduction band minimum ${\Gamma}_{\rm 1c}$ (valence band minimum ${\Gamma}_{\rm 8v}$) downwards (upwards). The present prediction of ${\Gamma}_{\rm 1c}$ and ${\Gamma}_{\rm 8v}$ as a function of the K doping reproduces well the widely tunable band gap, anisotropic Dirac semimetal state, and band-inverted semimetal state, as observed by angle-resolved photoemission spectroscopy experiment. Our findings shed new light on a route for tunable band gap engineering of 2D materials through the surface doping of alkali metals.
\end{abstract}

\maketitle

\section{INTRODUCTION}

Research on two-dimensional (2D) graphene has attracted enormous interests because of its unique massless Dirac fermion-like electronic structure, which provides a superior carrier mobility for potential applications in electronic devices~\cite{nov,nov2,zha}. However, such a gapless nature of graphene limits the on-off current ratio in field-effect transistors~\cite{xia}, hindering its utilization in electronic switching technologies. This inherent disadvantage of gapless graphene naturally leads to the development of new field-effect transistors using other 2D materials possessing intermediate band gaps, such as transition-metal dichalcogenides (TMDs)~\cite{dua,wang,tan}. It has been known that the band gaps of semiconducting TMDs can be tuned by varying film thickness~\cite{dua,wang}, allowing great flexibility in designing electronic and optical devices.

Similarly, as an elemental 2D material, black phosphorus (BP) also exhibits a tunability of band gap with respect to film thickness~\cite{liuh,kou,lil2}. Since monolayer or few-layer BP was successfully fabricated by mechanical exfoliation~\cite{liuh,lil}, the tunable band gap depending on film thickness has drawn much attention for its possible application in nanoelectronics~\cite{kou,ling,lil2}. Especially, monolayer BP has a puckered honeycomb lattice, giving rise to the armchair-shaped and zigzag-shaped patterns along the x and y directions, respectively [see Figs. 1(a) and 1(b)]. This anisotropic geometry of monolayer BP results in the interesting anisotropic thermal, optical, and electronic transport properties of few-layer BP~\cite{kou,ling,li}. Moreover, the band gap of monolayer or few-layer BP was found to be highly susceptible to strain and electric field~\cite{liu,rod,fei}. According to the density-functional theory (DFT) calculation of Liu $et$ $al$.~\cite{liu}, an externally applied electric field along the direction normal to the surface of few-layer BP shifts the valence and conduction bands upward and downward, respectively, representing the Stark effect. This electric field-induced Stark effect was found to produce the band inversion of valence and conduction bands with increasing the electric field, leading to the phase transition from a normal insulator to a topological insulator~\cite{liu} or a Dirac semimetal state~\cite{Baik,Ahn}. Meanwhile, by using the deposition of K atoms on the surface of BP, the angle-resolved photoemission spectroscopy (ARPES) experiment of Kim $et$ $al$.~\cite{kim} observed the widely tunable band gap, anisotropic Dirac semimetal state with a linear (quadratic) dispersion in the armchair (zigzag) direction, and band-inverted semimetal state. In order to simulate the low densities of K dopants using the DFT calculation, Kim $et$ $al$.~\cite{kim} artificially increased the vertical distance $d_{\rm v}$ between adsorbed K atoms and the four-layer BP surface from the optimized 2${\times}$2 structure (containing a single K atom in the 2${\times}$2 surface unit cell). From such a model calculation, they predicted a variation of band gap with respect to $d_{\rm v}$ and argued that the observed tunable band gap would be caused by a giant Stark effect induced by a vertical electric field from the positively ionized K overlayer to the negatively charged BP layers. This giant Stark effect in K-doped four-layer BP was treated in the same way as that~\cite{liu} earlier proposed in four-layer BP under the externally applied electric field across the film. However, the present study does not support such a underlying mechanism of the observed tunable band gap, as discussed below.

\begin{figure}[ht]
\centering{ \includegraphics[width=8.cm]{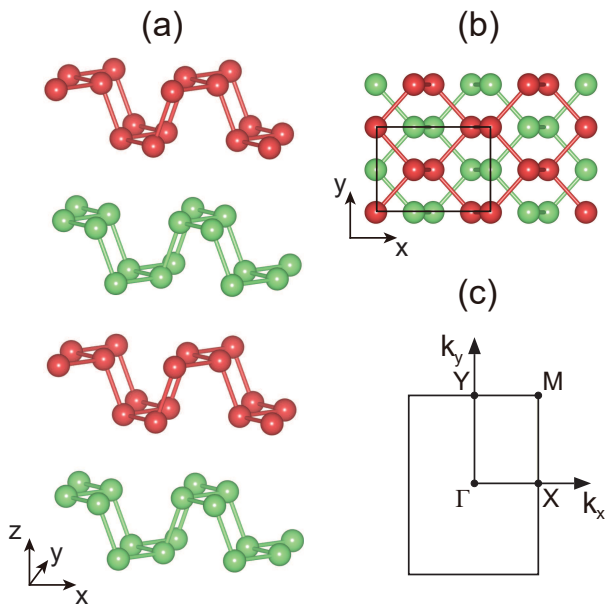} }
\caption{(a) Side view and (b) top view of the optimized structure of four-layer BP. For distinction, BP layers are alternatively drawn with red or green circles. In (b), the solid line represents the 1${\times}$1 unit cell. The surface Brillouin zone is given in (c).}
\end{figure}
Here, using first-principles DFT calculations, we investigate not only the adsorption and diffusion of the K atom on the surface of four-layer BP but also the underlying microscopic mechanism of the observed~\cite{kim} tunable band gap. We find that the K atom adsorbed on the BP surface has an energy barrier of ${\sim}$24 meV for the diffusion along the zigzag direction, which is much smaller than that (${\sim}$353 meV) along the armchair direction. This result indicates the easy (unfeasible) diffusion of K atoms along the zigzag (armchair) direction at the experimental~\cite{kim} temperature of 15$-$150 K. Such an anisotropic mobile feature of K atoms on the BP surface is most likely to produce the irregular adsorption patterns in the armchair direction depending on the in-situ deposition of K atoms in experiment~\cite{kim}. It is revealed that the donated electrons from K dopants are mostly localized at the topmost BP layer, which can efficiently screen the K ion potential. This extreme surface charging and its screening of K atoms determine the variations of the work function as well as the conduction and valence bands as a function of dopant density. We obtain two regimes in the variation of work function with increasing the K doping (equivalently electron doping): i.e., at the initial stage of doping, the work function is sharply decreased, while, as doping further increases, it is slowly increased. Meanwhile, the conduction band minimum ${\Gamma}_{\rm 1c}$ lowers monotonously relative to the Fermi level $E_{\rm F}$ with increasing electron doping, whereas the valence band maximum ${\Gamma}_{\rm 8v}$ lowers in the first regime but raises in the second regime, consistent with the ARPES data~\cite{kim}. These variations of ${\Gamma}_{\rm 1c}$ and ${\Gamma}_{\rm 8v}$ as a function of electron doping produce the widely tunable band gap, anisotropic Dirac semimetal state, and band-inverted semimetal state, as observed by ARPES experiment~\cite{kim}. The present findings provide a new microscopic picture for designing a widely tunable band gap of few-layer BP through the surface deposition of K atoms.

\section{COMPUTATIONAL METHODS}

Our DFT calculations were performed using the Vienna $ab$ $initio$ simulation package (VASP)~\cite{vasp1,vasp2} and the all-electron FHI-aims code~\cite{aims}. In the VASP, the projector-augmented wave potentials were employed to describe the interaction between ion cores and valence electrons~\cite{paw}. For the treatment of exchange-correlation energy, we used the generalized-gradient approximation functional of Perdew-Burke-Ernzerhof (PBE)~\cite{pbe}. The van der Waals (vdW) interactions were included using the PBE-D3 scheme~\cite{D3}. A plane wave basis was employed with a kinetic energy cutoff of 550 eV, and the ${\bf k}$-space integration was done with a 10${\times}$14 mesh in the surface Brillouin zone. The few-layer BP was modeled by a periodic slab geometry consisting of four layers with ${\sim}$20 {\AA} of vacuum in between the slabs. We employed a dipole correction that cancels the artificial electric field across the slab. All atoms were allowed to relax along the calculated forces until all the residual force components were less than 0.005 eV/{\AA}. For the simulation of electron doping, we used the virtual crystal approximation (VCA)~\cite{richter} implemented in the FHI-aims~\cite{aims} code to compensate excess electrons by varying the nuclear charge of the P atoms at the BP surface layer where the donated electrons from K dopants are mostly localized, as discussed below. The employed VCA scheme was successfully applied for
electron doping in the In/Si(111) surface system~\cite{sunwoo2015}.

\section{RESULTS}

\begin{figure*}[ht]
\centering{ \includegraphics[width=16.cm]{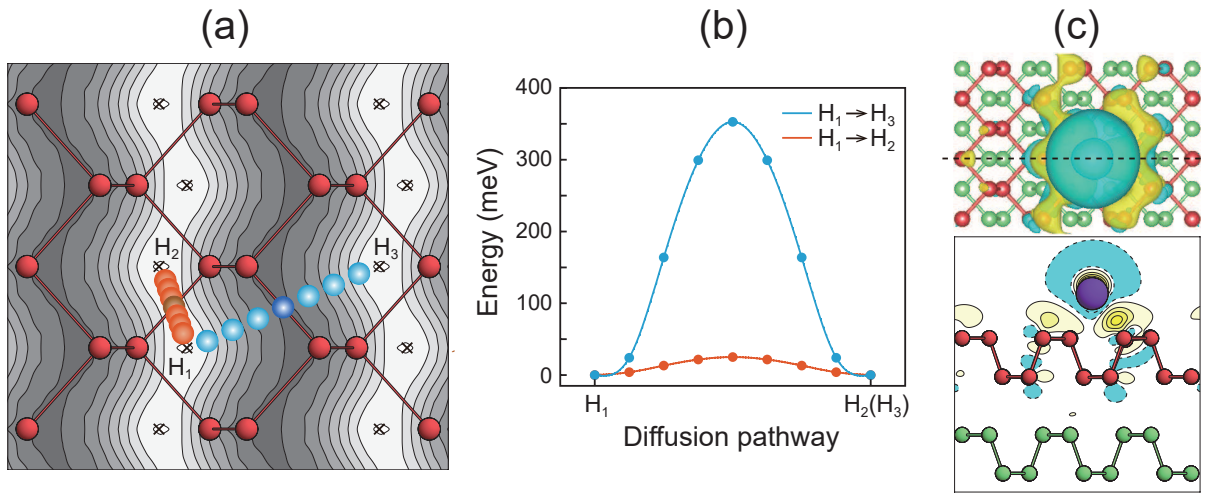} }
\caption{(a) Calculated PES for the adsorbed K atom on one BP surface. The contour spacing is 0.05 eV, and the energy zero is set to the H adsorption site (marked by ${\times}$). The minimum-energy diffusion pathways of K along the zigzag (H$_{\rm 1}$ $\rightarrow$ H$_{\rm 2}$) and armchair (H$_{\rm 1}$ $\rightarrow$ H$_{\rm 3}$) directions are drawn with circles. Here, the darker circle represents the transition state in each pathway. The corresponding energy profiles are displayed in (b). In (c), the calculated charge density difference ${\Delta}\rho$ (defined in text) is drawn with an isosurface of 0.5${\times}$10$^{-3}$ e/{\AA}$^3$. The contour plot in the vertical plane along the dashed line is also drawn with the first solid or dashed line of ${\pm}$0.3${\times}$10$^{-3}$ e/{\AA}$^3$ and a contour spacing of 1.5${\times}$10$^{-3}$ e/{\AA}$^3$. Here, the solid (dashed) line with yellow (cyan) color indicates accumulated (depleted) electrons.}
\end{figure*}

We begin to study the adsorption and diffusion of an isolated K atom on the BP surface using the PBE-D3 scheme. For this, we calculate the potential energy surface (PES) by optimizing the atomic structure of the K adatom within a large 3${\times}$3 supercell, which can make negligible the spurious interactions between K adatoms in the neighboring supercells. Figure 2(a) shows the contour plot of the PES, obtained from the 16${\times}$8 adsorption sites within a half of the 1${\times}$1 unit cell. We find that the K atom prefers to adsorb on the hollow (H) site with a binding energy ($E_{\rm b}$) of 1.89 eV. Here, $E_{\rm b}$ is defined as $E_{\rm tot}{\rm (BP)}$ + $E_{\rm tot}{\rm (K)}$ $-$ $E_{\rm tot}{\rm (K/BP)}$, where the first, second, and third terms are the total energies of the clean four-layer BP, the spin-polarized K atom, and the K adsorbed BP system, respectively. We also find that the diffusion energy barrier ($D_{\rm b}$) of K is ${\sim}$24 meV along the zigzag direction, which is significantly smaller than that (${\sim}$353 meV) along the armchair direction. Here, we use the nudged elastic band method~\cite{neb} to obtain $D_{\rm b}$ more accurately [see Fig. 2(b)]. Based on these results of the PES and $D_{\rm b}$, we estimate that (i) the K diffusion is kinetically more facilitated along the zigzag direction than the armchair direction by a factor of ${\sim}$10$^{11}$ at the experimental temperature of 150 K, and (ii) using an Arrhenius-type activation process with a typical pre-exponential factor of ${\sim}$10$^{13}$ Hz, the K diffusion rate from an H site to neighboring H site along the zigzag direction amounts to ${\sim}$8.6${\times}$10$^{4}$ (${\sim}$1.6${\times}$10$^{12}$) s$^{-1}$ at 15 (150) K. Therefore, we can say that, at the experimental~\cite{kim} temperature of 15$-$150 K, K atoms are restricted to easily diffuse along the zigzag direction, thereby giving rise to evenly distributed K adatoms at the H sites parallel to the zigzag chains. By artificially increasing $d_{\rm v}$ between the 2${\times}$2 K overlayer and the four-layer BP surface, Kim $et$ $al$.~\cite{kim} simulated the relatively weaker electric field at low dopant density. Since this DFT simulation of Kim $et$ $al$. gave a variation of the band gap $E_{\rm g}$ (= ${\Gamma}_{\rm 1c}$ $-$ ${\Gamma}_{\rm 8v}$) with respect to $d_{\rm v}$, they claimed that a vertical electric field from the positively ionized K donors modulates the band gap, owing to the giant Stark effect~\cite{kim}. However, the present study manifests that their prediction of $E_{\rm g}$ vs $d_{\rm v}$ is not due to the Stark effect, but can be attributed to the screened K ion potential, as discussed below.

Based on the calculated PES in Fig. 2(a), the transverse distribution of K atoms on the BP surface is likely to be determined upon their in-situ deposition because K adatoms are prevented from diffusing across the zigzag chains at the experimental~\cite{kim} temperature of 15$-$150 K. Despite such irregular adsorption patterns of K adatoms, the introduction of electron doping in few-layer BP can be accomplished through charge transfer from K dopants. Indeed, according to ARPES experiment~\cite{kim}, when the anisotropic Dirac band is formed with a band-gap closure, the amount of electron doping was experimentally estimated to be $n_c$ = 8.3${\times}$10$^{13}$ cm$^{-2}$, which is equivalent to ${\sim}$0.12 electrons (e) per 1${\times}$1 unit cell~\cite{coverage}. In order to estimate the electron transfer from adsorbed K atom to the BP layers, we calculate charge density difference defined as
\begin{equation}
{\Delta}\rho = {\rho}_{\rm K/BP} - ({\rho}_{\rm K} + {\rho}_{\rm BP}),
\end{equation}
where ${\rho}_{\rm K/BP}$ is the charge density of the equilibrium structure in Fig. 2(a) and ${\rho}_{\rm K}$+${\rho}_{\rm BP}$ is the superposition of the charge densities of the separated systems, i.e., the 3${\times}$3 K overlayer and the clean BP slab. Figure 2(c) shows that electron charge is transferred mainly from the region on top of the K atom into the region above the neighboring P atoms. It is thus likely that such a strong surface charging of donated electrons easily screens the electric potential of K ions.

To find the variation of $E_{\rm g}$ with respect to electron doping from K dopants, we calculate the band structure of four-layer BP with increasing excess electronic charge ($n$) per 1${\times}$1 unit cell. Such an electron doping into four-layer BP is simulated using the VCA~\cite{richter} without containing K adatoms on the BP surface. We note, however, that the screened potential of K ions is effectively incorporated in the VCA by increasing the nuclear charge of the P atoms at the BP surface layer. Figures 3(a)-3(d) show the calculated band structures of the clean and electron-doped ($n$ = 0.3, 0.37 and 0.4 e) four-layer BP, respectively. Here, we use the PBE calculation to obtain the geometry and band structure of each system~\cite{note1}. We find that the clean four-layer BP has $E_{\rm g}$ = 0.37 eV from the energy difference between ${\Gamma}_{\rm 1c}$ and ${\Gamma}_{\rm 8v}$ [see Fig. 3(a)]. This value of $E_{\rm g}$ is in good agreement with that (0.36 eV) of a previous PBE calculation~\cite{qiao}. It is noted that the ${\Gamma}_{\rm 8v}$ state has the partial densities of states (PDOS) of the $s$, $p_x$, $p_y$, and $p_z$ orbitals as a ratio of 7\%: 3\%: 0\%: 90\%. The most dominant $p_z$ orbital character of ${\Gamma}_{\rm 8v}$ represents an electron accumulation along the out-of-plane bonds of two P sublayers in the puckered honeycomb lattice [see Fig. 3(a)]. Meanwhile, the ${\Gamma}_{\rm 1c}$ state has the corresponding PDOS ratio as 18\%: 26\%: 0 \%: 56\%, indicating that the sum of $s$ and $p_x$ is comparable in magnitude with $p_z$. Consequently, contrasting with ${\Gamma}_{\rm 8v}$, the charge character of ${\Gamma}_{\rm 1c}$ shows not only nodes in the out-of-plane bonds but also some electron accumulation in the in-plane bonds within each P sublayer [see Fig. 3(a)].

\begin{figure*}[ht]
\includegraphics[width=14cm]{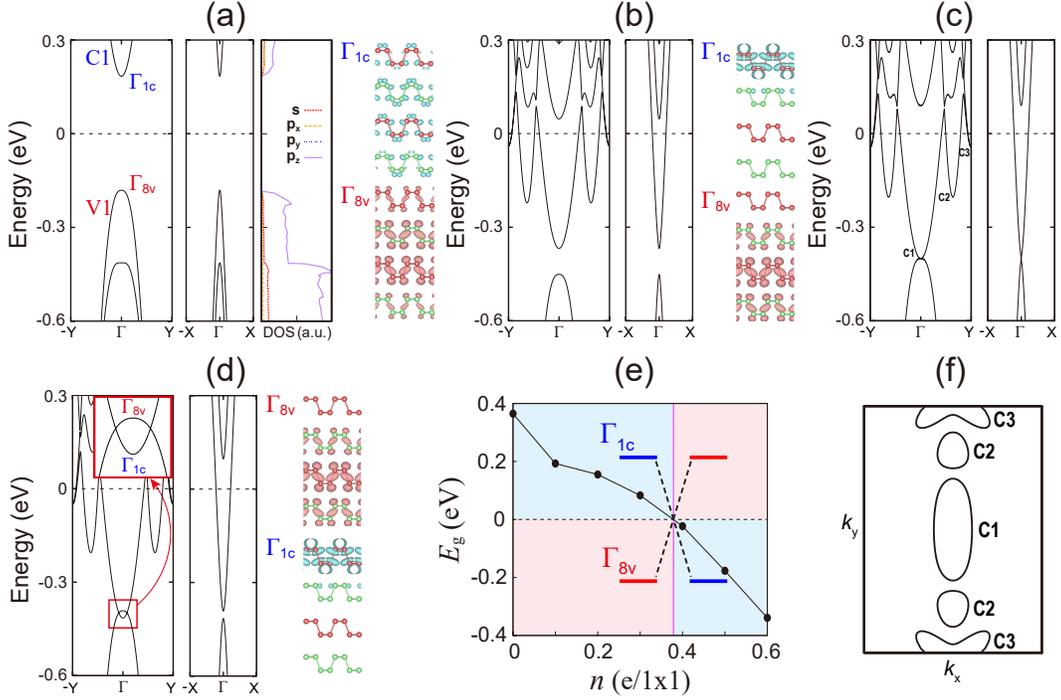}
\caption{Calculated band structures of four-layer BP using the VCA (a) without electron doping and with (b) $n$ = 0.3, (c) $n_c$ = 0.37 (the Dirac semimetal state), and (d) $n$ = 0.4 e per 1${\times}$1 unit cell. Here, ${\Gamma}_{\rm 1c}$ and ${\Gamma}_{\rm 8v}$ represent the conduction band (C1) minimum and the valence band (V1) maximum, respectively. The energy zero is set to the Fermi level. The charge characters of ${\Gamma}_{\rm 1c}$ and ${\Gamma}_{\rm 8v}$ are drawn with an isosurface of 0.5${\times}$10$^{-2}$ $e$/{\AA}$^3$. (e) $E_{\rm g}$ is displayed as a function of $n$ , where the inversion of the C1 and V1 bands is schematically given. The calculated Fermi surface of the Dirac semimetal state ($n_c$ = 0.37 e) within the surface Brillouin zone is given in (f).}
\end{figure*}

Figure 3(e) shows that $E_{\rm g}$ decreases monotonously with increasing $n$. We find that $E_{\rm g}$ closes at a critical value of $n_c$ = 0.37 e, leading to formation of a Dirac semimetal with the linear and parabolic band dispersions along the ${\Gamma}-$X and ${\Gamma}-$Y lines [see Fig. 3(c)], respectively. It is noted that electron doping changes the characters of ${\Gamma}_{\rm 1c}$ and ${\Gamma}_{\rm 8v}$ [see Figs. 3(b) and 3(d)]: i.e., ${\Gamma}_{\rm 1c}$ exhibits a significant localization at the surface layer, while ${\Gamma}_{\rm 8v}$ some localization below the surface layer. These characters of ${\Gamma}_{\rm 1c}$ and ${\Gamma}_{\rm 8v}$ are similar to those obtained under an external electric field applied normal to the surface~\cite{liu}. Specifically, for $n$ = 0.4 e, such charge characters of ${\Gamma}_{\rm 1c}$ and ${\Gamma}_{\rm 8v}$ clearly show that the conduction band C1 and the valence band V1 are inverted with each other around the ${\Gamma}$ point [see Figs. 3(d) and 3(e)]. This band inversion creates a pair of Dirac points with time-reversal symmetry along the ${\Gamma}-$Y line [see the inset of Fig. 3(d)]. These predictions of widely tunable band gap, anisotropic Dirac semimetal state, and band-inverted semimetal state with increasing $n$ are consistent with ARPES measurements~\cite{kim}. Although the present simulation does not take into account the electric field by excluding the presence of K adatoms above the surface, our results imply that the previously proposed~\cite{kim} giant Stark effect is not the main cause for the observed tunable band gap.

\begin{figure*}[ht]
\centering{ \includegraphics[width=16cm]{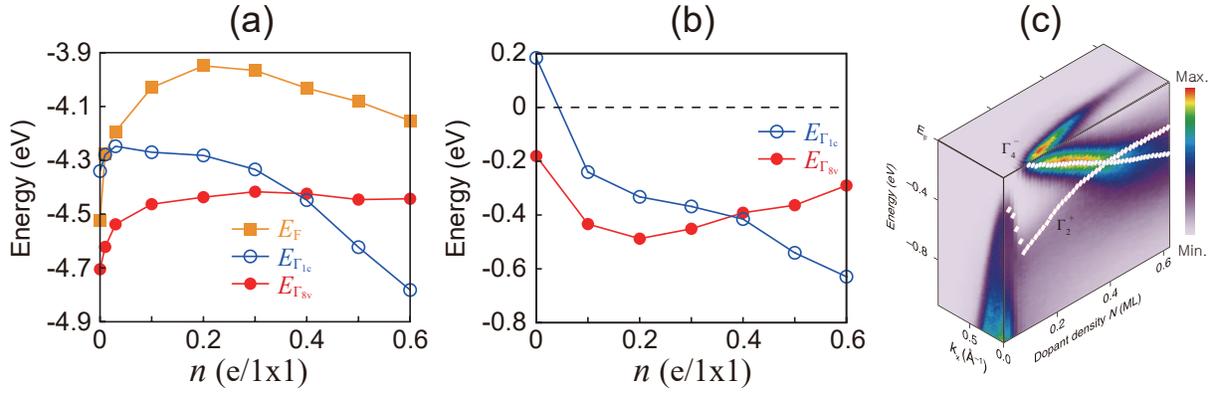} }
\caption{(a) Calculated $E_{\rm F}$, $E_{{\Gamma}_{\rm 1c}}$, and $E_{{\Gamma}_{\rm 8v}}$ relative to the vacuum level, as a function of $n$. In (b), $E_{{\Gamma}_{\rm 1c}}$ and $E_{{\Gamma}_{\rm 8v}}$ relative to $E_{\rm F}$ are also plotted. In (c), the ARPES data for the evolution of ${\Gamma}_{\rm 1c}$ (${\Gamma}_4^{-}$) and ${\Gamma}_{\rm 8v}$ (${\Gamma}_2^{+}$) as a function of K dopant density are taken from Science 349, 723 (2015). Reprinted with permission from AAAS. Here, blue circles and red diamonds denote $E_{{\Gamma}_{\rm 1c}}$ and $E_{{\Gamma}_{\rm 8v}}$ relative to $E_{\rm F}$, respectively.}
\end{figure*}

It is interesting to understand how the surface charging achieved by electron doping from K dopants produces the variation of $E_{\rm g}$. For this, we plot in Fig. 4(a) the Fermi level $E_{\rm F}$ and the binding energy $E_{{\Gamma}_{\rm 1c}}$ ($E_{{\Gamma}_{\rm 8v}}$) of ${\Gamma}_{\rm 1c}$ (${\Gamma}_{\rm 8v}$) as a function of $n$, relative to the vacuum level. It is seen that $E_{\rm F}$ raises sharply at the initial stage of doping between 0 ${\le}$ $n$ $<$ ${\sim}$0.2 e, but it lowers slowly when $n$ ${\ge}$ ${\sim}$0.2 e, giving rise to an initial decrease and a subsequent increase of the work function in the first and second regimes, respectively. Here, the upward shift of $E_{\rm F}$ in the first regime is due to the initial occupation of the C1 band, while the downward shift in the second regime is mostly attributed to the significant lowering of the C1 band [see Fig. 3(b)]. Interestingly, it is seen in Fig. 4(a) that, in the first regime, ${\Gamma}_{\rm 1c}$ slightly varies relative to the vacuum level, but it lowers sharply in the second regime. On the other hand, in the first regime, ${\Gamma}_{\rm 8v}$ shifts upward by 0.24 eV from the value obtained without electron doping, while it is nearly unchanged in the second regime [see Fig. 4(a)]. These contrasts of ${\Gamma}_{\rm 1c}$ and ${\Gamma}_{\rm 8v}$ under electron doping can be associated with the effect of the screened K ion potential due to surface charging: i.e., ${\Gamma}_{\rm 1c}$ possessing the localized charge character at the surface shifts toward high binding energy as $n$ increases, whereas ${\Gamma}_{\rm 8v}$ with a scarce surface charge is nearly insensitive with increasing $n$.

In Fig. 4(b), the variation of $E_{{\Gamma}_{\rm 1c}}$ and $E_{{\Gamma}_{\rm 8v}}$ relative to $E_{\rm F}$ is again plotted as a function of $n$. It is seen that ${\Gamma}_{\rm 1c}$ is monotonously lowered with increasing $n$, while ${\Gamma}_{\rm 8v}$ is initially lowered but subsequently raised. Overall, these variation patterns of ${\Gamma}_{\rm 1c}$ and ${\Gamma}_{\rm 8v}$ are consistent with the ARPES data~\cite{kim} which exhibits a monotonously downward shift of ${\Gamma}_{\rm 1c}$ and an initially downward but subsequently upward shift of ${\Gamma}_{\rm 8v}$ [see Fig. 4(c)]. Specifically, it is seen that, when $n$ ${\ge}$ ${\sim}$0.2 e, ${\Gamma}_{\rm 1c}$ and ${\Gamma}_{\rm 8v}$ are progressively closer to each other and eventually cross at $n_{\rm c}$ = 0.37 e. These downward and upward shifts of ${\Gamma}_{\rm 1c}$ and ${\Gamma}_{\rm 8v}$ with increasing $n$ were previously~\cite{kim} explained by the giant Stark effect which could be induced by the electric field from the positively ionized K atoms to the negatively charged BP layers. However, we here demonstrate that the upward shift of ${\Gamma}_{\rm 8v}$ is attributed to the above-mentioned lowering of $E_{\rm F}$ for n ${\ge}$ ${\sim}$0.2 e [see Fig. 4(a)], which is induced by the significant shift of ${\Gamma}_{\rm 1c}$ toward high binding energy due to the efficient screening of the K ion potential at the surface. We note that, if the electric field generated by the K overlayer is associated with the opposite shifts of ${\Gamma}_{\rm 1c}$ and ${\Gamma}_{\rm 8v}$, the adsorption of K atoms on both sides of the four-layer BP slab would not influence $E_{\rm g}$ because their created electric fields should cancel each other. However, our PBE calculation for such a structure of K adsorption with inversion symmetry shows that $E_{\rm g}$ still varies with increasing $d_{\rm v}$ (see Fig. 5), similar to the previous result~\cite{kim} obtained from the case of K adsorption on one side of the BP slab. Moreover, despite the fact that the in situ deposition of a certain amount of K atoms could yield irregular adsorption patterns in experiment (as discussed above), the ARPES data~\cite{kim} showed an invariance of the tunable
\begin{figure}[ht]
\centering{ \includegraphics[width=8.cm]{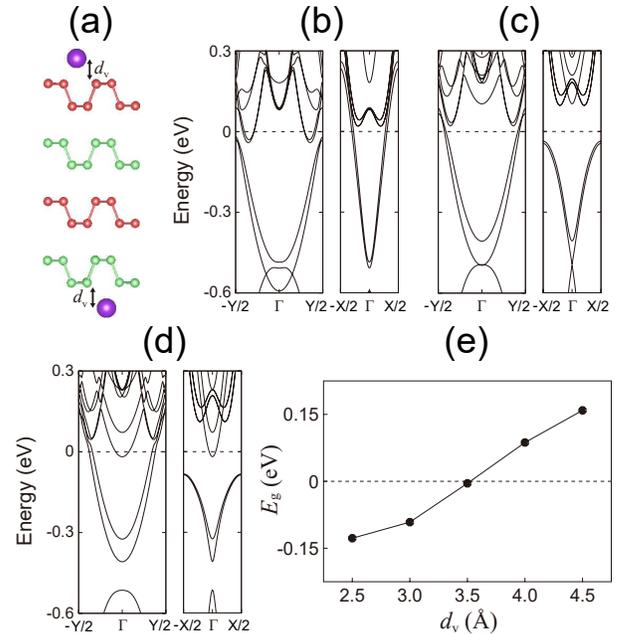} }
\caption{(a) Optimized 2${\times}$2 structure with K atoms on both sides of the four-layer BP slab and its band structures at (b) $d_{\rm v}$ = 2.5 {\AA} (the equilibrium state), (c) $d_{\rm v}$ = 3.5 {\AA}, and (d) $d_{\rm v}$ = 4.5 {\AA}. The energy zero is set to the Fermi level. In (e), $E_{\rm g}$ is displayed as a function of $d_{\rm v}$.}
\end{figure}
band gap as a function of dopant density~\cite{kim}. This also implies that the electric field from K adatoms to the BP layers would not play a major role in tuning $E_{\rm g}$. Thus, the previously proposed giant stark effect~\cite{kim} is unlikely the origin of the observed tunable band gap in K doped few-layer BP.

It is noteworthy that, although the present study captures essential features of how $E_{\rm g}$ is tuned via surface charging and its screening of K atoms, the predicted value of $n_{\rm c}$ = 0.37 e is much larger than that ($n_{\rm c}$ = 0.12 e) estimated from ARPES experiment~\cite{kim,coverage}. For this discrepancy, we discuss some reasons from the experimental and theoretical points of view. First, Kim $et$ $al$.~\cite{kim} may underestimate the value of $n_{\rm c}$, because they estimated the electron concentration by only taking into account the area of the C1 band enclosed by the Fermi surface using Luttinger's theorem~\cite{lutt}. However, according to our band structure obtained at $n_{\rm c}$, the Fermi surface encircles not only the C1 band around the ${\Gamma}$ point but also the C2 or C3 band along the positive and negative ${\Gamma}-$Y directions [see Figs. 3(c) and 3(f)]. Here, we obtain the ratio of their areas is C1 : C2 : C3 = 44.6\% : 24.6\% : 30.8\%. Using this ratio, we estimate that the experimental value of $n_{\rm c}$ can be increased to ${\sim}$0.27 e. Moreover, the effective mass is not a constant as the Luttinger's theorem assumes, but it increases as the electron filling increases, which further increases $n_{\rm c}$. It is noted that the series of ARPES intensity maps at constant energies showed strong suppression along the ${\Gamma}-$Y direction with respect to the ${\Gamma}-$X direction. Therefore, it is quite likely that the ARPES experiment of Kim $et$ $al$.~\cite{kim} could not detect the presence of the C2 and C3 bands. Secondly, unlike the experimental condition of rather thick BP samples~\cite{kim}, the four-layer BP slab employed in the present theoretical simulation with the PBE functional may induce some overestimation of $n_{\rm c}$. It is noted that doping induced band renormalization can also contribute to the band gap reduction~\cite{liang,suhuai}. Also, the presently employed VCA for electron doping may not accurately describe the screened K ion potential. In these respects, more refined experimental and theoretical works are anticipated in future to reduce the quantitative difference of $n_{\rm c}$ between the previous ARPES experiment~\cite{kim} and the present theory.

\vspace{0.4cm}
\section{SUMMARY}

We have performed a first-principles DFT study to explore the underlying mechanism of the observed tunable band gap in K doped few-layer BP. Unlike the previously proposed giant Stark effect~\cite{kim}, we demonstrated that the variation of $E_{\rm g}$ as a function of dopant density is caused by the ionic potential of K and the extreme surface charging of donated electrons and its screening of the K ion potential. It was found that, as electron doping increases, ${\Gamma}_{\rm 1c}$ monotonously lowers relative to the Fermi level, while ${\Gamma}_{\rm 8v}$ lowers at the initial stage of doping but raises subsequently. Consequently, we reproduced the widely tunable band gap, anisotropic Dirac semimetal state, and band-inverted semimetal state, in good agreement with the ARPES measurements~\cite{kim}. The present findings will offer a new aspect for designing a tunable band gap in 2D materials via surface charging with alkali-metal dopants.


\vspace{0.4cm}
\centerline{\bf ACKNOWLEDGEMENTS}
\vspace{0.4cm}

This work was supported by National Research Foundation of Korea (NRF) grant funded by the Korean Government (Nos. 2015M3D1A1070639 and 2015R1A2A2A01003248). S.W is supported
by the NSFC (No. U1530401). The calculations were performed by KISTI supercomputing center through the strategic support program (KSC-2016-C3-0059) for the supercomputing application research. S.W.K. acknowledges support from POSCO TJ Park Foundation.

\noindent Corresponding authors: $^{\dagger}$suhuaiwei@csrc.ac.cn, $^{*}$chojh@hanyang.ac.kr


\begin{thebibliography}{99}
\bibitem{note1} Using the PBE-D3 calculation, we obtained $E_{\rm g}$ = 0.09 eV for a clean four-layer BP, in good agreement with that (0.07 eV) of a previous PBE-D2 calculation~\cite{liu}. Since this value of $E_{\rm g}$ is largely underestimated compared to that (${\sim}$0.3 eV) measured by ARPES experiment~\cite{kim}, we performed the PBE calculation which properly predicted the magnitude of $E_{\rm g}$.
\end{thebibliography}
\end{document}